\documentclass[conference]{IEEEtran}

\usepackage{graphicx} 
\usepackage{refstyle}
\usepackage[font=footnotesize]{caption}
\usepackage{flushend}

\usepackage{float} 
\usepackage{wrapfig} 
\usepackage{amsmath}
\usepackage{lipsum} 

\graphicspath{{Pictures/}} 

\usepackage{amsmath}

\newtheorem{prop}{Proposition}

\newtheorem{definition}{Definition}

\newcommand{\be}{\begin{eqnarray}}
\newcommand{\ee}{\end{eqnarray}}
\newcommand{\beeq}{\begin{equation}}
\newcommand{\eeeq}{\end{equation}}
\newcommand{\beeqs}{\begin{eqnarray*}}
\newcommand{\eeqs}{\end{eqnarray*}}

\begin{document}
%
\title{KERMAN: A Key Establishment Algorithm based on Harvesting Randomness in MANETs}

\author{\IEEEauthorblockN{Mohammad Reza Khalili Shoja\IEEEauthorrefmark{1},
George Traian Amariucai\IEEEauthorrefmark{1},
Shuangqing Wei\IEEEauthorrefmark{2} and 
Jing Deng\IEEEauthorrefmark{3}}
\IEEEauthorblockA{\IEEEauthorrefmark{1}Department of Electrical and Computer Engineering, Iowa State University, mkhalili@iastate.edu, gamari@iastate.edu}
\IEEEauthorblockA{\IEEEauthorrefmark{2}School of Electrical Engineering and Computer Science, Louisiana State University, swei@lsu.edu}
\IEEEauthorblockA{\IEEEauthorrefmark{3}Department of Computer Science of University, North Carolina at Greensboro, jing.deng@uncg.edu}}
\maketitle

\makeatletter{\renewcommand*{\@makefnmark}{}
\footnotetext{This material is based upon work supported in part by the National Science Foundation under Grant No. 1320351.}\makeatother}

\begin{abstract}
Establishing secret common randomness between two or multiple devices in a network resides at
the root of communication security. The problem is traditionally decomposed into a randomness
generation stage (randomness purity is subject to employing often costly true random number generators)
and a key-agreement information exchange stage, which can rely on public-key infrastructure or on
key wrapping. In this paper, we propose KERMAN, an alternative key establishment algorithm for ad-hoc
networks which works by harvesting randomness directly from the network routing metadata, thus
achieving both pure randomness generation and (implicitly) secret-key agreement. Our algorithm
relies on the route discovery phase of an ad-hoc network employing the Dynamic Source Routing
protocol, is lightweight, and requires minimal communication overhead.
\end{abstract}
\IEEEpeerreviewmaketitle


\section{Introduction}\label{intro}

Automatic key establishment between two devices in a network is generally performed either by public-key-based
algorithms (like Diffie-Hellman \cite{diffie1976new}), or by encrypting the newly-generated key with a special
\emph{key-wrapping key} \cite{Bellare00}. However, in addition to the well-established, well-investigated keying
information exchange, one additional aspect of key establishment is often understated: to ensure the security
of the application it serves, the newly generated secret key has to be truly random. While minimum standards
for software-based randomness quality are generally being enforced \cite{park1988random}, many applications
rely on often costly hardware-based \emph{true random generators} \cite{sunar2009true}. Sources of randomness
employed by true random number generators vary from wireless receivers and simple resistors to ring oscillators and SRAM
memory.

In this paper, we build upon the observation that a readily-available source of randomness is usually neglected:
the network dynamics. Indeed, by their very nature, networks are dynamic and largely unpredictable. Their
randomness is usually evident in easily-accessible networking metadata such as traffic loads, packet delays or
dropped-packet rates. However, as the main focus of our work is on mobile ad-hoc networks (MANETs), the source
of randomness we shall discuss in this paper is one that is specific to infrastructure-less networks:
the routing information itself. Another interesting feature of the routing information, in addition to its
randomness, is that it can easily be made available to the devices that took part in the routing process, but
it is usually unavailable to those devices that were not part of the route. This idea opens the door to a whole new
class of applications: with the proper routing protocol, the routing information could be used for establishing
\emph{secret common randomness} between any two devices in a mobile ad-hoc network. This common randomness could then
be further processed into true common randomness, and used as secret keys.

Common randomness was pioneered in \cite{maurer.93.IT, ahlswede.93.IT, ahlswede.98.IT.part2}, where it is shown that
if two parties, Alice and Bob, have access to two correlated random variables (RVs) $X'$ and $Y'$ respectively, (in either
the source or the channel models), a secret key can be established between them through public discussions and random-binning-like
(e.g. hashing) operations. The key should remain secret from an adversary eavesdropper (Eve) who overhears the public discussions,
and possesses side information (in the form of a third RV $Z$) correlated with that available at Alice and Bob. Common-randomness-based
key establishment generally consists of three phases. First, Alice and Bob have to agree on two other RVs $X$ and $Y$, such that
$H(X|Y)<H(X|Z)$ and $H(Y|X)<H(Y|Z)$, where $H(\cdot)$ is the standard Shannon entropy. This part is sometimes called \emph{advantage
distillation}. Next, Alice and Bob (and also Eve)  sample their respective random variables a large number of times, producing
sequences of values. Then Alice and Bob exchange further messages (over a public channel) to agree on the same single
sequence of values -- this phase is the \emph{information reconciliation}. Finally, because the agreed-upon sequence is not
completely unknown to Eve (Eve can sample her variable $Z$ synchronously with Alice and Bob), Alice and Bob run a randomness
extractor on it, to produce a secret key (a shorter sequence) which, from Eve's perspective, is uniformly distributed over its
space -- this is the \emph{privacy amplification phase}. The ideas of \cite{maurer.93.IT, ahlswede.93.IT} have been recently
applied to secret key generation in wireless systems, where secure \emph{common randomness} is attained by exploiting reciprocal
properties of wireless channels or other auxiliary random sources in the physical layer \cite{Wallace.2010, bloch.2008, ye.mathur.2010,
agrawal.rezki.2011, wang.infocom.2011,ren.2011.mag, T.chou.IT.2012, Ye.IT.2012, Khisti.IT.2012}.
One noteworthy observation is that, while the work of \cite{maurer.93.IT, ahlswede.93.IT, ahlswede.98.IT.part2} considers
an information-theoretic approach, in practice Alice and Bob do not usually have access to large numbers of values drawn from
their random variables, but rather to only one or a few values. To address this issue, \cite{Renner1} shows that for such single-shot
scenarios, the smooth minimum entropy gives a tight upper bound on the achievable size of the secret key.

In MANETs, the lack of infrastructure, the nodes' mobility and the fact that packets are routed by nodes, instead of fixed
devices, have resulted in the need for specialized routing protocols, like the ad-hoc on-demand distance vector AODV routing, or
the dynamic source routing (DSR) \cite{RFC-DSR}. For our secret-common-randomness-extraction purposes, DSR appears to be the
ideal candidate, and will be the object of this paper. DSR contains two main mechanisms -- Route Discovery and Route Maintenance --
which work together to establish and maintain routes from senders to receivers. The protocol works with the use of explicit
\emph{source routing}, which means that the ordered list of nodes through which a packet will pass is included in the packet header.
It is sets of these routing lists that we shall show how to process into secret keys shared between pairs of nodes.
 
Our contributions can be summarized as follows:
\begin{enumerate}
\item We show that the randomness inherent in an ad-hoc network can be harvested and used for establishing secret keys between
pairs of nodes that participate in the routing process.
\item We provide a very practical algorithm for establishing such secret common randomness, based on the DSR protocol, and
we demonstrate a way to calculate an achievable number of shared secret bits, using an adversary's beliefs.
\item We simulate a realistic ad-hoc network in OPNET Modeler, and show that within only ten minutes, thousands of secret
bits can be shared between different node pairs.
\item We discuss methods to optimize the DSR-based key establishment process between two nodes -- using the spoiling
knowledge technique and implementing an optimized set partition algorithm.
\end{enumerate}

The rest of this paper is organized as follows. Those parts of the DSR protocol that are essential for understanding our algorithm
are examined in Section \ref{dsr}. In Section \ref{model}, we describe the system model and state our assumptions.
Section \ref{kerman} describes our proposed key establishment algorithm. Simulation results obtained with OPNET Modeler are presented
and discussed in Section \ref{simulation}, while Section \ref{conclusion} draws conclusions and discusses future work.

\section{Dynamic Source Routing}\label{dsr}

Dynamic source routing (DSR) \cite{RFC-DSR} is one of the well-established routing algorithms for ad-hoc networks.
Under this protocol, when a user (the sender) decides to send a data packet to a destination, the sender must insert the \emph{source route}
in a special position of the packet's header, called the \emph{DSR source route option}. The \emph{source route} is an ordered list of nodes
that will help relay the packet from its source to its destination. The sender transmits the packet to the first node in the \emph{source route}.
If a node receives a packet for which it is not the final destination, the node will transmit the packet to the next hop indicated by
the \emph{source route}, and this process will continue until the packet reaches its destination.

To obtain a suitable source route toward the destination, a sender first searches its own \emph{route cache}. The \emph{route cache}
is updated every time a node learns a new valid path through the network (whether or not the node is the source or the destination for that path).
If no route is found after searching the route cache, the sender initiates the \emph{route discovery} protocol. During the route discovery,
the source and destination become the \emph{initiator} and \emph{target}, respectively. 

\begin{figure}
\center{\includegraphics[scale=.75]{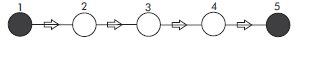}}
\caption{Communication among node 1 and 5}
\label{fig:DSR}
\end{figure}
As a concrete example, suppose node 1 in \figref{DSR} wants to send packets to node 5. Initially, node 1 does not have any route toward node 5,
and thus node 1 initiates a route discovery by transmitting a single special local broadcast packet called \emph{route request}.
The \emph{route request option} is inserted in the packet's header, following the IP header. To send the route request, the source address
of the IP header must be set to the address of the initiator (node 1), while the destination address of IP header must be set to the IP
limited broadcast address. These fields must not be changed by the intermediate nodes processing the route request.
A node initiating a new route request generates a new identification value for the route request, and places it in the ID field of the route
request header. The route request header also contains the address of the initiator and that of the target. The route request ID is meant
to differentiate between different requests with the same initiator and target -- it should be noted here that the same request may reach
an intermediate or destination node twice, over different paths.
Each route request header also contains a record listing the address of each intermediate node through which this particular copy of the route
request has been forwarded. In our example, the route record initially lists only the address of the initiator node 1. As the packet reaches node 2,
this node inserts its own address in the packet's route record, and broadcasts it further, and so on, until the packet reaches the target node 5,
at which point its route record contains a valid route (1-2-3-4-5) for transmitting data from node 1 to node 5.

As a general rule, recent route requests received at a node should be recorded in the node's \emph{route request table} -- the sufficient information
for identifying each request is the tuple (initiator address, target address, route request ID). When a node receives a route request packet, several
scenarios can occur. First, if the node has recently seen another route request message from this initiator, carrying the same id and target address,
or if this node’s own address already exists in the route record section of the route request packet, this node discards the route request.
Second, if the route request is new, and if the node is the target, it sends a \emph{route reply} packet to the initiator, and saves a copy of the
route (extracted from the route request route record) in a table called the \emph{route cache}. Third, if the request is new, but the node is not the
target, the node inserts its address in the packet's route record, and broadcasts the modified packet.

In our example in \figref{DSR}, node 5 constructs  a \emph{route reply} packet and transmits it to the initiator of the route request (node 1).
The source address of the IP header of the route reply packet is set to the IP address of sender of route reply (node 5). In our example,
node 5 is also the target. But this need not occur. Under the DSR protocol, it is possible that an intermediate node (who is not the target
of the route request) already has a path to the target in its route cache. Then it is this node that transmits the route reply back to the
initiator, and it is its IP address that gets inserted in the source IP address part of the route reply packet's header. The route reply packet header
also contains a \emph{route record}. This route record starts with the address of the first hop after the initiator and ends with the address of
the target node (regardless of whether the node that issues the route reply is the target or not). In our example,
the route record contained in the route reply packet is (2, 3, 4, 5). Including the address of the initiator node 1 in the route record
would be redundant, as the address of node 1 is already included as the destination address in the IP header of the route reply packet.
The combination of the route record and destination address in the IP header is the \emph{source route} which the initiator will use for
reaching its target. It is also noteworthy that network routes are not always bidirectional. That is, it may not always be possible for
node 5 to send its route reply to node 1 using a route obtained by simply inverting the source route. In the more general case,
node 5 has to search its own route cache for a route back to node 1. If no such path is found, node 5 should perform its own route
discovery for finding a \emph{source route} to node 1.

\section{System Model}\label{model}

Mobile ad-hoc networks (MANETs) consist of mobile nodes communicating wirelessly with each other, without any pre-existing
infrastructure. We consider a \emph{bidirectional} MANET employing dynamic source routing (DSR), in which the nodes
(corresponding to the mobile devices of the network's users) are moving in a random fashion in a pre-defined area.
The bidirectional network assumption is usually a practical one, especially when all the nodes in the network belong to
the same class of devices (e.g. smart phones)\footnote{It should be noted that our algorithm should work (albeit with some reduction
in performance) even if the network is not bidirectional. In this case, the route request ID needs to be inserted in the route
reply packet. The reduction in performance for this scenario follows from the security considerations -- namely, more nodes are
involved in the routing mechanism, and hence have access to the source route.}.

According to the route discovery protocol outlined in section \ref{dsr},
every single node in the network is assumed equally likely to be the initiator of a route request packet, at any given time.
Furthermore, we assume that the target of any route request is uniformly distributed among the remaining nodes.
Any route discovery instance will return a path through the network (the source route), of a given length. The length
of a returned path is distributed according to a probability distribution that depends on all the parameters of the network.
Deriving a model for this probability distribution, based on the network parameters, is outside the scope of this work. Hence,
in the remainder of this paper, we shall assume that all nodes have access to such an (empirically-derived) probability
distribution over the path lengths. That is, if we denote the random variable describing the length of some path $r$ by
$L_r$, then we assume that all the nodes have access to the prior $p(L_r=l)$, for $l=2,3,\ldots$. For our experiments,
we run our simulation for a long time, and derive $p(L_r=l)$ by counting the paths of equal length.
We also assume that \emph{all paths of the same length are equally probable}. To express this notion, denote the random variable that samples
a path (or a partial path) by $R$. Then we can write $p(R=r|L_r=l)=\frac{1}{N_l}$ if the length of path $r$ is $l$ (otherwise
the probability is zero), where $N_l$ is the total number of paths of length $l$. This leads to $p(R=r)=\frac{1}{N_{l_r}}p(L_r=l_r)$,
where $l_r$ is the actual length of path $r$.

Our protocol, called \emph{KERMAN} runs by making each node collect in a table all the source routes that it is part of -- recall
that since the network is assumed to be bidirectional, a node can extract the route request ID, the initiator and the target from the route
request packet, save them in a temporary table, and then, if a route reply packet carrying a source route with the same initiator and target
is observed within a pre-determined time interval, the node can associate the source route with the route request ID, and save both in a
long-term table.

This mechanism brings about our security model. Since the common randomness established between two nodes by our algorithm consists of
the source routes, it should be clear that several other nodes can be privy to this information. For instance, all the nodes included
in a particular source route have full knowledge of this route. Moreover, it is likely that the route reply packet carrying a source
route can be overheard by malicious eavesdroppers that are not part of the source route at all. Therefore, to achieve a level of
security, two nodes will have to gather a large collection of source routes, such that \emph{none of the other nodes that appear in any
of the source routes in this collection has access to all the routes in the collection}. Unfortunately this is not enough, because
it is still possible that one of the nodes, most likely a node that is part of many -- though not all -- routes in the collection,
eavesdropped on all the remaining routes that it is not part of.

We deal with this problem by making an additional assumption: we assume that any two source routes are exchanged under independent
and uniformly distributed network arrangements. That is, for the exchange (route discovery) of each source route, all the nodes in
the network are distributed uniformly, and independently of other exchanges, in their pre-defined area. Moreover, the network remains
the same for the entire duration of the route discovery and the associated data transmission. These assumptions are
appropriate when the network nodes move around fast relative to the time between two different route discovery phases, but slow
relative to the duration of a single communication session.
This means that for any source route, the probability that any node which is \emph{not} itself part of the route overhears the route
(by overhearing a route reply or a data packet) is only a function of the network parameters. In the remainder of this section, we
show how to compute the probability that an eavesdropper Eve knows a source route of which it is not part.

Denote the binary random variable encoding whether an eavesdropper Eve overhears a source route $r$ by $K_{Eve}(r)$.
Then $p(K_{Eve}(r)=1)$ depends on: (a) Eve's reception radius, (b) the total area of the network (all the places where
Eve could be during the communication session corresponding to source route $r$), and (c) the length of the path.
\begin{figure}
\center{\includegraphics[scale=.75]{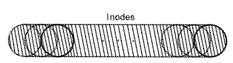}}
\caption{The area covered by l nodes }
\label{fig:Area1}
\end{figure}
The computation is described in \figref{Area1}, where it can be observed that the worst-case scenario for a path of length $l$
is when all the $l$ nodes are arranged in a straight line. In this case, we can use the following worst-case approximation
(obtained by first calculating the area of a circular segment):
\be\label{pkgivenl}
p(K_{Eve}(r)=1|L_r=l)=\nonumber\\
=\frac{l\pi d_e^2-2(l-1)d_e^2(\frac{\pi}{3}-\frac{\sqrt{3}}{4})}{S_{total}}=\frac{d_e^2(1.91 \cdot l +1.23)}{S_{total}},
\ee
where $d_e$ is the maximum eavesdropping range (the radius of the circles in \figref{Area1}), which is assumed the same for
each of the nodes (all nodes transmit with the same power, using isotropic antennas), and $S_{total}$ is the total area
of the pre-defined location where the nodes can move.

Finally, two additional assumptions are made: the attackers are purely passive eavesdroppers (as attackers -- otherwise,
they are allowed to initiate well-behaved communication, just like any other node), and they do not collude.


\section{Proposed Algorithm }\label{kerman}

In this section we introduce KERMAN, a \emph{K}ey-\emph{E}stablishment algorithm based on \emph{R}andomness harvested from the source routes
in a \emph{MAN}ET employing the DSR algorithm. To establish secret common randomness between two nodes in the MANET, KERMAN uses the
standard sequence of three steps outlined in Section \ref{intro}: advantage distillation, information reconciliation and privacy amplification.
\begin{figure}
\center{\includegraphics[scale=.75]{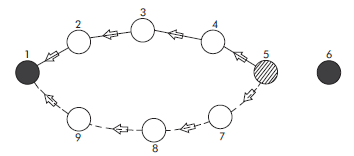}}
\centering
\caption{Example for proposed algorithm}
\label{fig:myprotocol1}
\end{figure}

\subsection{Advantage Distillation} \label{advdistill}
To accomplish advantage distillation, every node in the network has to maintain a new table called the \emph{ Selected Route Table}, or SRT.
The SRT contains source routes that include that node's address. To demonstrate how the SRT is built, we consider the following example.
Take the scenario in \figref{myprotocol1}, in which node 1 and 6 are the source and the destination, respectively. Since node 1 does not
have any route to node 6, it generates and broadcasts a route request packet. Assume that the id of this packet is 14, which means that this
is the fourteenth attempt that node 1 makes to reach node 6. As seen in \figref{myprotocol1}, node 5 will generate the route reply from its
own route cache (because we assumed that node 5 already knows how to reach node 6). The transmission path of the route reply from node 5 to
node 1 has been illustrated in \figref {myprotocol1}, and is consistent with a bidirectional network. Each intermediate node that receives
this route reply inserts the source route in their own SRT. The SRT has three columns dubbed \emph{RID}, \emph{partial route} and \emph{full route}
respectively. RID is a tuple that consists \emph{(Source IP, Destination IP, route request ID, route-reply-sender IP)}. In our scenario, nodes
1, 2, 3, 4 and 5 will all record an entry in their respective SRTs, with the RID 1-6-14-5. The intermediate nodes (2, 3 and 4) can obtain the
route request ID by searching their own \emph{route request tables} as discussed in Section \ref{dsr}. The \emph{partial route} field
of the SRT entry identifies those other nodes that are supposed to have this particular route in their SRT -- in this case, nodes 1, 2, 3, 4 and 5.
The \emph{full route} field is the entire route from source to destination, which will be used for data transmission (1,2,3,4,5,6 in this case).
The SRTs of the nodes 1, 2, 3, 4 and 5 have the same following entry:
\begin{center}
\footnotesize
    \begin{tabular}{| l | l | l |}
    \hline
    RID & Partial Route & Full Route\\ \hline
	1-6-14-5 & 1-2-3-4-5 & 1-2-3-4-5-6\\
\hline
    \end{tabular}
\end{center}
It should be noted that, because node 6 did not directly hear the route request from node 1, it has no way of determining the route request ID in the
RID, and this is why it cannot store this entry in its SRT, although it will most likely learn the source route from the received data packets that
follow the route discovery phase. Regardless, when discussing the security of the established secret common randomness, node 6 will be assumed to have
full knowledge of the \emph{full route}.

Each full route in a nodes' SRT is only available to a limited number of nodes in the network, i.e., those nodes which are included in in the
source (full) route, along with some nodes who are not part of the source route but happen to overhear the route request and route reply exchange.   
The following proposition states that SRT entries are unique in the whole network.
 \begin{prop}
If two nodes have the same RID in their own SRTs, then the full routes associated with this RID in two SRTs are exactly the same.
\end{prop}
\begin{proof}
Based on the DSR protocol \cite{RFC-DSR}, in the phase of processing a received route request, several steps must be performed in a well-defined order.
The step consisting of the search in the route request table is done before the phase of sending route reply from the route cache. But if, while searching
the route request table, a node finds that it has received this route request before, the node must discard the route request packet. Hence, an intermediate
node can initiate the route reply only in response to the first route request, and will ignore all subsequent route requests with the same ID, source and
destination. Therefore, it is not possible that multiple route replies originate from the same node in response to the same route request, even if the route
request was received multiple times, via different paths. Now, although two different route replies in response to the same route request can originate
at different nodes, (for example, in \figref{myprotocol1} node 7 also knows a path to node 6 and initiates a route reply), the RIDs corresponding to
these route replies contain the IP of the route reply sender, and hence are different.
\end{proof}

\subsection{Information Reconciliation}
Information reconciliation is usually a complex process, involving techniques from channel or source coding, and displaying very restrictive
lower bounds on the amount of information that needs to be transmitted over a public channel \cite{Renner1} -- these bounds can often leave
very little uncertainty for an eavesdropper. Fortunately, KERMAN is particularly well-suited for information reconciliation, and only requires
minimal communication overhead.

Let us assume that two nodes --call them Alice and Bob for simplicity -- realize that they share a large number of routes in their SRTs. For
instance, Alice could first notice that Bob is part of a large number of partial routes in her SRT, and could ask Bob to perform information
reconciliation, with the purpose of eventually generating a shared secret key. Upon Bob's acceptance, Alice sends him the list of RIDs
corresponding to the partial routes in Alice's SRT that include the address of Bob. Bob can then verify whether he already has the received
RIDs in his SRT, and can send back to Alice only those RIDs that he could not locate. The information reconciliation is now complete.
Alice and Bob share a set of full routes, which constitute their common randomness.

There is but one caveat. As mentioned in Section \ref{advdistill}, the RIDs consist of the tuples (Source IP, Destination IP,
route request ID, route-reply-sender IP) corresponding to each route request/ route reply pair. Moreover, it is possible
that Alice and Bob are neither the source nor the destination, nor the route-reply sender. Thus, transmitting an RID in the
clear, over a public channel, may expose up to five nodes of the route (source, destination, route-reply sender, Alice and Bob)
to an eavesdropping adversary. While this does not prevent KERMAN from working, many of the full routes of length less than six
would become useless for our purposes. An alternative option is to only transmit a (non-keyed) hash of each RID. If the hash
is pre-image resistant, and in addition its output is short (although long enough to ensure that no collision takes place
for those entries in Bob's SRT which contain Alice's address in the full route field), the information leaked to an adversary
Eve would be much less than that leaked from the entire RID.

The following simplifying assumptions make the treatment in the current version of this paper more tractable, without restricting
the applicability of our results. Removing the assumptions is a straightforward extension, but requires further discussion of
random hash functions (i) and a sophisticated probabilistic argument (ii). (i) We assume that, when transmitting
the RIDs from Alice to Bob, no information leaks about the contents of the RIDs, except the addresses of Alice and Bob. (ii) We
assume that every node in the network can see that Alice and Bob exchange RIDs and has access to the information allowed in (i).

\subsection{Privacy Amplification}
For the purposes of this section we shall represent the full routes as sets of node identifiers, or addresses. Alice and Bob share a list of common
full routes. Now Alice and Bob can construct the set $\mathcal{M}=\{m_1,m_2, \ldots, m_h\}$ where $m_i$ (we'll call it a \emph{trimmed route}) is
produced from the full route $r_i$, by removing the addresses of Alice and Bob. At this point, full routes and trimmed routes are in a one-to-one
correspondence. However, for following the remainder of this section, it is essential that the reader remembers the difference between a full
route and a trimmed route.

In the next step, Alice partitions the set of trimmed routes $\mathcal{M}$ into several \emph{disjoint} subsets $\mathcal{M}_k\subset mathcal{M}$ of
various sizes $h_k$, such that, for any $\mathcal{M}_k=\{m_{1,k},m_{2,k},\ldots,m_{h_k,k}\}$, the probability that any node in the network
has knowledge of all the $h_k$ trimmed routes is less than a small security parameter $\epsilon_1$. This means that, with probability larger than
$1-\epsilon_1$, there exists at least one trimmed route in $\mathcal{H}$ that Eve knows nothing about -- note that this is true for any identity
that Eve may take (except, of course Eve cannot be Alice or Bob). It is the full route corresponding to this trimmed route (different from any
node's perspective) that constitutes the randomness of the generated secret.

To extract a secret from each of the sets $\mathcal{M}_k$, Alice first represents all the \emph{full routes} by pre-determined binary strings of
the same length. The length of the strings is determined as the logarithm to base two of the total number of possible full routes, in a
practical scenario. For example, from our simulations, we noticed that full routes are limited to 15 nodes, which means that trimmed routes
are limited to 13 nodes. In a network of 50 nodes, there are thus $\binom{48}{1}3!+\binom{48}{2}4!+\ldots +\binom{48}{13}15!$ possible full
routes involving Alice and Bob, where the factorial terms account for all the possible arrangements. For example, there are $\binom{48}{1}$
trimmed routes of length 1, and their corresponding full routes have length 3 (this includes the unknown node that defines the trimmed route,
Alice and Bob), and there are $3!=6$ possible arrangements of these three nodes. This total number of possible full routes amounts to
representing each full route on 78 bits. The binary sequences representing the \emph{full routes} corresponding to the trimmed routes in
$\mathcal{M}_k$ are then XORed together.
The result is inserted into a $(k,\epsilon_2)$-randomness extractor \cite{Shaltiel} (defined below), which outputs a shorter bit string
$s_k$ -- the secret. The secret $s_k$ should satisfy the $(\epsilon_1,\epsilon_2)$-security defined below. The following series of
definitions -- some of them included for the sake of completeness -- formalizes this procedure.

\begin{definition}
\emph{Minimum Entropy: From \cite{Renner1}:}
Let X be a random variable with alphabet $\mathcal{X}$ and probability distribution of $P_X (x)$. The min-entropy is defined as
\begin{equation} 
H_{\infty}(X)=  -log \, \max\limits_{x \in \mathcal{X}} \, P_X (x).  
\end{equation}
\end{definition}

\begin{definition}
\emph{Smooth Minimum Entropy: From \cite{Renner1}:}
Let X be a random variable with alphabet $\mathcal{X}$ and probability distribution of $P_X (x)$, and let $\epsilon_3 > 0$. The $\epsilon_3$-smooth
min-entropy of $X$ is defined as
\begin{equation} 
H_{\infty}^{\epsilon_3}(X)=  -log \, \max\limits_{\mathcal{Q}_{X} \in \mathcal{B}^{\epsilon_3}(P_{X})} \, Q_{X} (x)    
\end{equation}
where the maximum ranges over the $\epsilon_3$-ball $\mathcal{B}^{\epsilon_3}(P_{X})$ \cite{Renner1}.
\end{definition}

\begin{definition}
\emph{Extractor: From \cite{Shaltiel}:}
A function E: $\{0,1\}^n \times \{0,1\}^d\to \{0,1\}^s$ is a $(k,\epsilon_2)$-extractor if for every distribution $X$ over $\{0,1\}^n$,
with $H_{\infty}(X)\geq k, E(X, Y)$ is $\epsilon_2$-close to uniform, where $Y$ is distributed uniform and it is independent of $X$.
Here $\epsilon_2$-close refers to the \emph{statistical distance}\cite{Shaltiel}.
\end{definition}

Randomness extractors are functions that generate almost uniform bits (truly random bits) from weak random sources. Note that
the extractor defined above is a seeded extractor, i.e. it requires a few additional bits of (non-secret) randomness.

\begin{definition}
In the context of a MANET, a piece of secret common randomness $s_k$ established between two nodes Alice and Bob is called
$(\epsilon_1, \epsilon_2)$-secure if, with probability larger than $1-\epsilon_1$, the secret $s_k$ is $\epsilon_2$-close to
uniform from the perspective of any node in the network, except Alice and Bob.
\end{definition}

It has been shown in \cite{Renner1} that the number of completely random bits that can be extracted from a bit sequence
should be upper bounded by, but very close to, the smooth min-entropy of the sequence. Thus, for the purposes of this paper,
we shall only focus on the (smooth) minimum entropy of a full route, viewed from the perspective of an eavesdropper who does not
know anything about the associated trimmed route. This minimum entropy is a good indication of the number of secret random
bits that can be extracted from each set $\mathcal{M}_k$.
In the remainder of this section we shall show how to calculate, how to smooth out (if possible), and how to artificially increase
this minimum entropy.

Calculating this min-entropy is equivalent to computing a probability distribution that characterizes Eve's belief about
the unknown full route. The task is not straightforward because of two reasons. First routes of different lengths have different
probabilities of appearing in SRTs (these depend on the network parameters). Second, if a route is longer, then the probability that
Eve has accidentally overheard it is larger -- recall (\ref{pkgivenl}). Therefore, we start off with an empirically-derived prior
$p(L_r=l_r)$ denoting the probability that the unknown route has length $l_r$, and with
$p(K_{Eve}(r)=0|L_r=l_r)=1-p(K_{Eve}(r)=1|L_r=l_r)$ from (\ref{pkgivenl}), and we compute
\be\label{plgivenk}
p(L_r=l_r|K_{Eve}(r)=0)=\nonumber\\
=\frac{p(L_r=l_r) p(K_{Eve}(r)=0|L_r=l_r)}{\sum_{\substack{l}} p(L_r=l) p(K_{Eve}(r)=0|L_r=l)}.
\ee
As explained above, from Eve's perspective there are $\binom{N-2}{l_r-2}l_r!$ equally-likely full routes of length $l_r$
and containing Alice and Bob, where $N$ is the total number of nodes in the MANET. Thus we can write the probability that
the unknown full route is $r$ (where $r$ has the length of $l_r$) as:
\be\label{proute}
p(r|K_{Eve}(r)=0))=\frac{p(L_r=l_r|K_{Eve}(r)=0)}{\binom{N-2}{l_r-2}l_r!},
\ee

\section{Simulation Results}\label{simulation}

\subsection{OPNET Simulation and Results}\label{opnet}

The proposed protocol has been simulated in OPNET Modeler, using the parameters indicated in table \ref{table:nonlin}.    
This choice of parameters results in a maximum eavesdropping range of $d_e=12$m.
\begin{table}[ht]
\caption{Simulation Parameters} 
\centering 
\begin{tabular}{|c|c|} 
\hline 
Simulation Parameters & Value\\[1ex] 
\hline 
Network Size & 100m*100m  \\ 
Number of Nodes & 50  \\
Simulation Duration & 600(sec) \\
Transmit Power(w) & .005  \\
Packet Reception-Power Threshold(dBm) & -55 \\  
Speed(meters/seconds) & uniform(.5,1)\\
Packet Inter-Arrival Time(seconds) & exponential(1)\\
\hline 
\end{tabular}
\label{table:nonlin} 
\end{table}

\begin{figure}
\center{\includegraphics[scale=.5]{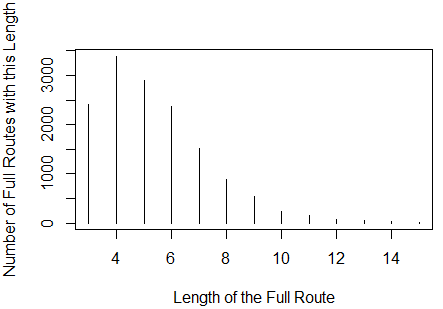}}
\caption{Number Of Full Routes vs. Full Route Length}
\label{fig:result2}
\end{figure}

\begin{figure}
\center{\includegraphics[scale=.5]{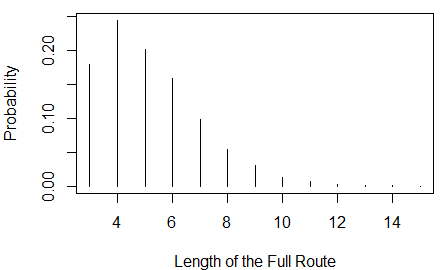}}
\caption{Posterior Probability Distribution of an Unknown Route's Length}
\label{fig:prob4}
\end{figure}

Each node sends packets to four random destinations.
The number of full routes vs the full route length is shown in \figref{result2}, and
the empirically-derived prior $p(L_r=l_r)$ looks similar.
The posterior probability $p(L_r=l_r|K_{Eve}(r)=0)$ is illustrated in \figref{prob4}. 

To calculate the min entropy, we need the probability distribution of the unknown full route. According to (\ref{proute}), this is
$p(r|K_{Eve}(r)=0))=\frac{p(L_r=l_r|K_{Eve}(r)=0)}{\binom{48}{l_r-2}l_r!}$. Due to the obtained values being more than 20 orders of magnitude apart, we chose to
represent $p(r|K_{Eve}(r)=0))$ in table \ref{table:aa}.
\begin{table*}
\caption{Probability Distribution of an Unknown Full Route, from Eve's Perspective}
\label{table:aa}
\begin{tabular*}{\textwidth}{@{\extracolsep{\stretch{1}}}*{14}{|c}| @{}}
 \hline
Length Of the Full Route & 3 & 4 & 5 & 6 & 7 & 8 & 9 & 10 & 11 & 12 & 13 & 14 &15    \\
\hline
probability & 0.00062	& 9.02E-06	& 9.7E-08	& 1.1E-09	& 1.1E-11 & 1.1E-13&1.2E-15	& 9.9E-18&1E-19	&1E-21	&1.6E-23	& 2E-25 &2E-27 \\
      \hline                       
\end{tabular*}
\end{table*}

It can be easily seen that $H_{min}(r|K_{Eve}(r)=0))=-\log_2(0.00062)=10.66$, and that smoothing out the probability distribution, with
a sensibly small smoothing parameter $\epsilon_3$ cannot increase the min entropy. In other words, about 10 bits can be extracted from each
subset $\mathcal{M}_k$ of full routes. Now the remaining question is how many subsets $\mathcal{M}_k$ we can form. To solve this problem,
for any pair of nodes we organize the full set of all trimmed routes $\mathcal{M}$ as a  \emph{selection matrix}.
In the \emph{selection matrix}, a row corresponds to one of the trimmed routes in $\mathcal{M}$. A column corresponds to a node's address.
There are 38 columns (one for each node in the MANET, except Alice and Bob). Each entry in the matrix is the probability that the node
in the respective column knows the full route corresponding to the respective row. The selection matrix can be represented as follows:
$$\bordermatrix{\text{}&node~1&node~2&\ldots &node~t\cr
                m_1&a_{11} &  a_{12}  & \ldots & a_{1t}\cr
                m_2& a_{21}  &  a_{22} & \ldots & a_{2t}\cr
               \vdots  & \vdots & \vdots & \ddots & \vdots\cr
                m_h& a_{n1}  &   a_{n2}       &\ldots & a_{nt}}$$
where $a_{ij}$ is the probability that node $j$ knows full route $i$. For example, when node $j$ is a part of the full route corresponding
to the trimmed route $i$, then  $a_{ij}=1$. Otherwise, $a_{ij}=p(K_{j}(i)=1|L_i=l_i)$, where $l_r$ is the length of the full route $i$.
Now the partitioning algorithm consists of constructing \emph{distinct} sub-matrices $\mathcal{M}_k$, each consisting of $h_k$ rows of
$\mathcal{M}$, such that the product of the entries in each column of $\mathcal{M}_k$ be less than $\epsilon_1$. We shall informally
call this property $\epsilon_1$-security. An optimal partition maximizes the number of sub-matrices $\mathcal{M}_k$ with the $\epsilon_1$-security property.

For this paper, we chose to implement a na\"{\i}ve partition algorithm. In this algorithm, Alice builds $\mathcal{M}_1$ by selecting the first row in the
selection matrix, and adding the next row in the selection matrix, until the column-wise product condition holds. Then Alice moves
to the next row, and starts building $\mathcal{M}_2$, and so on, until she runs out of columns in $\mathcal{M}_k$. For two different $\epsilon_1$, the number of subsets in the whole
network is shown in table \ref{table:ab}.  
\begin{table}
\caption{Number of subsets, obtained by the na\"{\i}ve algorithm, when considering all full routes of length at least 3.
Total network-wide achievable number of shared secret bits, in last column.}
\label{table:ab}
\begin{tabular}{|c|c|c|c|c|c|c|c||c|}
 \hline
No. of Subsets & 1 & 2 & 3 & 4 & 5 & 6 & 7 & $B_{total}$ \\
\hline
No. of Pairs, $\epsilon_1=10^{-3}$ & 215 & 75 & 22 & 6 & 1 & 0 & 1 & $4.98\cdot 10^3$ \\
\hline
No. of Pairs, $\epsilon_1=10^{-4}$ & 192& 58& 7& 3& 1 & 1& 0 & $3.75\cdot 10^3$\\
\hline                       
\end{tabular}
\vspace{-4mm}
\end{table}
To provide a basis for comparison with the results in the next subsection, we also calculate the maximum achievable total network-wide
number of shared random bits (between all the possible pairs in the network), $B_{total}$ -- this is shown in the last column of
table \ref{table:ab}. For example, for $\epsilon_1=10^{-3}$, we have $B_{total}=10.66\cdot(215\cdot 1+75\cdot 2+22\cdot 3+6\cdot 4+1\cdot 5+1\cdot 7)$.

\subsection{Increasing The Secret's Length by Spoiling Knowledge}

Spoiling knowledge was introduced in \cite{bennett1995generalized} as a means of (publicly) adjusting a probability distribution
to increase its min entropy. In our specific example, this translates to purposely discarding the most likely full routes from the SRT.
But since all routes of the same length have the same probability (from Eve's perspective), we can only increase the min entropy
by discarding all the routes of a given length. The downside, of course, is that the number of partitions satisfying the properties
outlined in Section \ref{opnet} also decreases.

For our specific scenario, disregarding the full routes of length 3 yields a min entropy of roughly $H_{min}(r|K_{Eve}(r)=0))=-\log_2(9.02\cdot 10^{-6})=16.76$
bits. For two different values of $\epsilon_1$, the number of subsets in the whole network is shown in table \ref{table:ac}.
Again, the last column of the table shows the maximum achievable total network-wide number of shared random bits $B_{total}$.
It is interesting to note that for $\epsilon_1=10^{-3}$ and $\epsilon_1=10^{-4}$, the spoiling knowledge technique
achieves a gain of roughly 23\% and 15\%, respectively.

\begin{table}
\caption{Number of subsets, obtained by the na\"{\i}ve algorithm, when considering only full routes of length at least 4.
Total network-wide achievable number of shared secret bits, in last column.}
\label{table:ac}
\begin{tabular}{|c|c|c|c|c|c|c|c||c|}
 \hline
No. of Subsets & 1 & 2 & 3 & 4 & 5 & 6 & 7 & $B_{total}$\\
\hline
No. of Pairs, $\epsilon_1=10^{-3}$ & 195 & 60 & 11 & 2 & 2 & 0 & 0 & $6.13\cdot 10^{3}$ \\
\hline
No. of Pairs, $\epsilon_1=10^{-4}$ & 159& 37& 5& 0& 2 & 0& 0 & $4.32\cdot 10^{3}$\\
\hline                       
\end{tabular}
\vspace{-4mm}
\end{table}

\section{Conclusions and Future Work}\label{conclusion}

We have shown that the randomness inherent in an ad-hoc network can be harvested and used for establishing
shared secret keys. For practical network parameters, we have demonstrated that after only ten minutes of
use, thousands of shared secret bits can be established between various pairs of nodes. This number can
be further increased by the spoiling knowledge technique of \cite{bennett1995generalized}. While we showed
how this works at the entire-network level, a better option might be to let each one of the pairs of nodes
decide whether using the spoiling knowledge technique is advantageous or not.

The number of achievable shared secret bits can be further increased by devising a more efficient partition
algorithm for the generation of full-route subsets with the $\epsilon_1$-security property, instead of the
na\"{\i}ve algorithm used in this paper.

Finally, although this paper focuses on the routing information circulated by DSR, other types of randomness
can be exploited, such as the network's connectivity or traffic load. These untapped sources of
randomness are the subject of future work.

\bibliographystyle{IEEEtran}
\bibliography{Mycitation,lsuisuunc}
\end{document}